\documentclass[journal]{IEEEtran}

\usepackage{cite}

\ifCLASSINFOpdf
\usepackage[pdftex]{graphicx}

\else

\fi

\usepackage{amsmath}
\usepackage{amssymb}
\usepackage{amsfonts}
\usepackage{amsthm}
\usepackage{verbatim}
\usepackage{algorithmic}
\usepackage{algorithm}

\usepackage{array}
\allowdisplaybreaks[4]
\begin{document}
%
\onecolumn
\title{Communication-Efficient Local SGD with \\
Age-Based Worker Selection}

\author{Feng~Zhu,
        Jingjing~Zhang,~\IEEEmembership{Member,~IEEE}
        and~Xin~Wang,~\IEEEmembership{Senior~Member,~IEEE}
\thanks{The authors are with the Key Laboratory for Information Science of Electromagnetic Waves (MoE), Department of Communication Science and Engineering, Fudan University, Shanghai 200433, China (e-mail: \{20210720072, jingjingzhang, xwang11\}@fudan.edu.cn).)

}} 
\maketitle

\begin{abstract}
A major bottleneck of distributed learning under parameter-server (PS) framework is communication cost due to frequent bidirectional transmissions between the PS and workers. To address this issue, local stochastic gradient descent (SGD) and worker selection have been exploited by reducing the communication frequency and the number of participating workers at each round, respectively. However, partial participation can be detrimental to convergence rate, especially for heterogeneous local datasets. In this paper, to improve communication efficiency and speed up the training process, we develop a novel worker selection strategy named AgeSel. The key enabler of AgeSel is utilization of the ages of workers to balance their participation frequencies. The convergence of local SGD with the proposed age-based partial worker participation is rigorously established. Simulation results demonstrate that the proposed AgeSel strategy can significantly reduce the number of training rounds needed to achieve a targeted accuracy, as well as the communication cost. The influence of the algorithm hyper-parameter is also explored to manifest the benefit of age-based worker selection.
\end{abstract}


\IEEEpeerreviewmaketitle

\section{Introduction}
\IEEEPARstart{P}{arameter-Server (PS)} setting is one of the most popular paradigms in distributed machine learning. In this setting, the PS broadcasts the current global model parameter to the workers for gradient computation and aggregates the computed gradients to update the global model. The operation repeats until some targeted convergence criterion is reached\cite{dean2012large,li2014scaling,lian2017can,yan2022performance}. However, the massive communication overhead between the PS and the workers has become the main bottleneck of the overall system performance as the sizes of the neural networks grow exponentially enormous\cite{bottou2010large}. 

In order to address the communication issue, \cite{mcmahan2017communication} proposes a widely-studied algorithm, named federated averaging (FedAvg), where the PS randomly selects a subset of workers and sends the global model to the selected workers for a certain number of local updates at each round. The PS then collects the latest local models to update the global model and re-sends it to the re-selected workers for local updates again. 
By decreasing the communication frequency and the number of participating workers, FedAvg can achieve improved communication efficiency, with its convergence analysis extensively studied under both homegenous and heterogeneous data distribution cases in e.g., \cite{li2019convergence,yang2020achieving,zhou2018convergence,haddadpour2019local,wang2019adaptive2,yu2019parallel}. Additionally, \cite{uddin2021robust} proposes a robust aggregation scheme to improve the performance of FedAvg. 

To further improve communication efficiency, instead of random selection as in FedAvg, adaptive worker selection technique has attracted increasing attention recently. One of the representative works on adaptive selection in SGD-based distributed learning is \cite{chen2021cada}, in which a worker is required to upload its gradient only if its contribution (i.e., the change to the global model) is large enough or its local gradient has become overly stale. The asserted optimal client selection strategy (OCS) is further developed in \cite{chen2020optimal} by selecting the workers with larger gradient norms to minimize the variance of the global update. In addition, another metric of 
“contribution”, i.e., local loss of workers, is also explored as a criterion for worker selection design \cite{goetz2019active,ribero2020communication,cho2020bandit}.




On the other hand, imbalanced participation could render the training unstable or slow, especially under heterogeneous data distribution, since the influence of some parts of the data on the overall training process is weakened \cite{yin2018gradient,kaul2012real}.
Hence, the age of each worker, indicating the number of consecutive rounds where it has not participated in the computation, could be taken into consideration by adaptive selection techniques. To this end, \cite{ozfatura2020age} explores the influence of ages in gradient descent (GD) scenario. For distributed learning over wireless connections, \cite{9053740} jointly leverages the channel quality and the ages of workers to improve the communication efficiency. Additionally, if with perfect lossless channel, the scheme reduces to the Round Robin policy. Moreover, \cite{buyukates2021timely} finds the age-optimal number of workers to select at each round, where the age is defined as the sum of computation time and uplink transmission time.


In this paper, we propose a novel age-based worker selection strategy for local SGD. In what we call AgeSel scheme, a simple age-based mechanism is implemented such that the workers are forced to be selected if they have not been involved for a certain number of consecutive rounds. Different from existing adaptive worker selection strategies, the proposed AgeSel relies on the age information that is readily available at the PS without additional communication and computation overhead. The convergence of AgeSel is also rigorously established to justify the benefit of age-based partial worker participation. The simulation results corroborate the superiority of AgeSel in terms of communication efficiency and required number of training rounds (to achieve a targeted accuracy) over state-of-the-art schemes. 


\textbf{Notations.} $\mathbb{R}$ denotes the real number fields; $\mathbb{E}$ denotes the expectation operator; $\|\cdot\|$ denotes the $\ell_2$ norm; $\nabla F$ denotes the gradient of function $F$; $\bigcup$ denotes the union of sets; $\mathcal{A}\subseteq\mathcal{B}$ represents that set $\mathcal{A}$ is a subset of set $\mathcal{B}$; and $|\mathcal{A}|$ denotes the size of set $\mathcal{A}$.

\section{Problem Formulation}
\subsection{System Model}
Consider the PS-based framework of distributed learning with heterogeneous data distribution. There are $M$ distributed workers in set $\mathcal{M}\triangleq\{1,...,M\}$. Each worker $m$ maintains a local dataset $\mathcal{D}_m$ of size $N_m$. It is drawn from the global dataset $\mathcal{D}=\{z_i\}_{i=1}^N$, i.e., we have $\mathcal{D}=\bigcup_{m\in\mathcal{M}} \mathcal{D}_m$. Our objective is to minimize the weighted loss function
\begin{align}
    \mathcal{L}(\boldsymbol{\theta})=\sum_{m=1}^M\frac{N_m}{N} \mathcal{L}_m(\boldsymbol{\theta}), \label{objective}
\end{align}
where $\boldsymbol{\theta}\in\mathbb{R}^d$ is the $d$-dimension parameter to be optimized, and we have the definition $\mathcal{L}_m(\boldsymbol{\theta})\triangleq\mathbb{E}[\mathcal{L}_m(\boldsymbol{\theta};z_m)]$, with $\mathcal{L}_m(\boldsymbol{\theta})$ being the local loss function of worker $m$ and $z_m$ being the sample drawn randomly from its local dataset $\mathcal{D}_m$.

To elaborate on local SGD, we define $\boldsymbol{\theta}^j$ as the global model parameter at training round $j$, and $\boldsymbol{\theta}_m^{j,0}$ as the local model of worker $m$ before operating local updates. At each training round $j$, a subset $\mathcal{M}_D^j\subseteq\mathcal{M}$ of workers are  selected to download the global model $\boldsymbol{\theta}^j$ from the PS prior to computation. After that, each selected worker $m$ in $\mathcal{M}_D^j$ sets its local model as $\boldsymbol{\theta}_m^{j,0}=\boldsymbol{\theta}^j$ and starts operating $U$ local iterations, with the updating formula given as
\begin{align}
    \boldsymbol{\theta}_{m}^{j,u+1}=\boldsymbol{\theta}_{m}^{j,u}-\frac{\eta}{B} \sum_{b=1}^B \nabla \mathcal{L}(\boldsymbol{\theta}_{m}^{j,u}; z_{m,b}^{j,u}), \label{update}
\end{align}
for any local iteration $u=0,...,U-1$. Here $\eta$ is the stepsize and $\frac{1}{B} \sum_{b=1}^B \nabla \mathcal{L}(\boldsymbol{\theta}_{m}^{j,u}; z_{m,b}^{j,u})$ is the minibatch gradient to be computed by worker $m$ at local iteration $u$, where $B$ is the minibatch size and $z_{m,b}^{j,u}$ is a sample drawn independently across iterations from the local dataset $\mathcal{D}_m$.

After all the workers in set $\mathcal{M}_D^j$ have completed their local computations, a subset of $S$ workers $\mathcal{M}_U^j\subseteq\mathcal{M}_D^j$ are selected to upload their latest models, and the PS aggregates the models
\begin{align}
    \boldsymbol{\theta}^{j+1}=\frac{1}{S}\sum_{m\in\mathcal{M}_U^j}\boldsymbol{\theta}_m^{j,U}.\label{aggregation}
\end{align}
Note that here we employ the unbiased aggregation since the weight of each worker is reflected in the worker selection process, as will be specified later.

The training process ends when some stopping criterion is satisfied, with the total number of rounds denoted as $J$.





\subsection{Performance Metrics}
To gauge the efficiency of the proposed scheme, we are interested in two performance metrics, i.e., the number of training rounds and the communication cost required to reach a targeted training accuracy.

Firstly, the number of training rounds $J$ of the algorithm to achieve a targeted test accuracy is used to reflect the training speed, as well as the communication cost of the algorithm.

Secondly, we define the communication cost $C^j$ at training round $j$ as the number of communication rounds between the PS and the workers, given as 
\begin{align}
    C^j=|\mathcal{M}_D^j|+S.\label{comm cost}
\end{align}
This is due to the fact that the size of the global parameter
downloaded by the workers is the same as that of the latest
parameter uploaded by each selected worker. As a result, the total communication cost is given as $C=\sum_{j=0}^{J-1} C^j$.

These two metrics will be used in Section IV to compare the performance of different schemes.

\section{Adaptive Selection in Local SGD (AgeSel)}

In this section, we first develop the novel AgeSel scheme that aims at improving communication efficiency by utilizing the age-based worker selection. Then the convergence of AgeSel is rigorously established.

\subsection{Algorithm Description}
To elaborate on worker selection strategy in the proposed AgeSel, we define an $M$-length vector $\boldsymbol{\tau}_M$ to collect the ages of the workers, as in \cite{chen2021cada,chen2020optimal}. Note that $\boldsymbol{\tau}_M$ is maintained by the PS and initialized to be a zero vector. Each element $\tau_m$ of vector $\boldsymbol{\tau}_M$ measures the number of consecutive rounds that worker $m$ has not been selected by the PS. Particularly, at each round $j$, $\tau_m$ is updated as follows:
\begin{equation}
    \tau_m=
    \begin{cases}
      0, ~~~~~~~~\text{if $m$ is selected},  \\
      \tau_m+1, ~\text{otherwise}.
    \end{cases}
\end{equation}
To identify the workers with low participation frequency, we pre-define a threshold $\tau_{max}$. Accordingly, worker $m$ would be forced to participate when having its $\tau_m\geq \tau_{max}$. The specific procedure of AgeSel is delineated next.


 
At each round $j$, the PS selects $S$ workers to perform computation by checking the vector $\boldsymbol{\tau}_M$. More precisely, with vector $\boldsymbol{\tau}_M$, we can identify all the infrequent workers with their ages greater than $\tau_{max}$, which would be considered first. Let $S^j$ denote the number of these workers. The set $\mathcal{M}_D^{j}$ of selected workers can be determined as follows. For the first case with $S^j\geq S$, the PS simply picks the $S$ workers in an age-descending order. Note that the the workers with larger sizes of local datasets are prioritized when there are ties in ages. For the second case with $S^j < S$, the PS first picks all the $S^j$ infrequent workers and then chooses the rest $S-S^j$ without replacement from the workers having ages smaller than $\tau_{max}$ with the probabilities proportional to the sizes of their datasets. 


Once the set $\mathcal{M}_D^{j}$ is determined, the PS then broadcasts the global parameter $\boldsymbol{\theta}^j$ to all the selected workers. By initiating its local model with $\boldsymbol{\theta}_m^{j,0}=\boldsymbol{\theta}^j$, each worker $m$ in $\mathcal{M}_D^j$ then starts $U$ iterations of local updates through (\ref{update}) and sends its latest model $\boldsymbol{\theta}_m^{j,U}$ to the PS, i.e., we have $\mathcal{M}_U^j=\mathcal{M}_D^j$. Eventually, the PS updates the vector $\boldsymbol{\tau}_M$, along with the global model aggregated via (\ref{aggregation}).

Note that when $\tau_{max}$ is very large, we barely have infrequent workers. Then the set $\mathcal{M}_D^{j}$ is indeed chosen by weights (i.e., the sizes of datasets), as with the FedAvg algorithm\cite{mcmahan2017communication}; i.e., in this case, AgeSel is reduced to FedAvg.

\textit{Merits of AgeSel:} The benefit from the age-based mechanism used in AgeSel is two-fold. First, it has been shown that less participation of some workers can be detrimental to convergence rate due to the lack of gradient diversity \cite{yin2018gradient}, especially for heterogeneous local datasets. Our age-based selection strategy can balance the worker participation, thereby preserving gradient diversity and ensuring the fast convergence. Second, generation of age information $\boldsymbol{\tau}_M$ here incurs no extra communication and computation cost, in contrast to other information such as the (costly) norm of updates \cite{chen2020optimal}, used in existing alternatives. Hence, AgeSel is more communication- and computation-efficient.


\subsection{Convergence Analysis}
We next establish the convergence of the proposed AgeSel algorithm under heterogeneous data distribution, with a general (not necessarily convex) objective function. Our analysis is based on the following two assumptions, which are widely adopted in related works such as \cite{li2019convergence, yang2020achieving}.

\textbf{Assumption 1} (Smoothness and Lower Boundedness) \textit{Each local 
function $\mathcal{L}_m(\boldsymbol{\theta})$ is $L$-smooth, i.e.,}
\begin{align}
    &\left\|\nabla \mathcal{L}_m(\boldsymbol{\theta}_1)-\nabla \mathcal{L}_m(\boldsymbol{\theta}_2)\right\| \leq L \left\|\boldsymbol{\theta}_1-\boldsymbol{\theta}_2\right\|,\label{lipschitz_gradient}
\end{align}
\textit{$\forall \boldsymbol{\theta}_1, \boldsymbol{\theta}_2\in \mathbb{R}^d$. We also assume that the objective function $\mathcal{L}$ is bounded below by $\mathcal{L}^*$.}

\textbf{Assumption 2} (Unbiasedness and Bounded Variance) \textit{For the given model parameter $\boldsymbol{\theta}$, the local gradient estimator is unbiased, i.e.,}
\begin{align}
    \mathbb{E}[\nabla\mathcal{L}_m(\boldsymbol{\theta};z)]=\nabla\mathcal{L}_m(\boldsymbol{\theta}).
\end{align}
\textit{Moreover, both the variance of the local gradient estimator and the variance of the local gradient from the global one are bounded, i.e., there exist two constants $\sigma_L, \sigma_G>0$, such that}
\begin{align}
    &\mathbb{E}[\|\nabla\mathcal{L}_m(\boldsymbol{\theta};z)-\nabla\mathcal{L}_m(\boldsymbol{\theta})\|^2]\leq \sigma_L^2, \forall m\\
    &\mathbb{E}[\|\nabla\mathcal{L}_m(\boldsymbol{\theta})-\nabla\mathcal{L}(\boldsymbol{\theta})\|^2]\leq \sigma_G^2, \forall m.
\end{align}


With these assumptions, we can derive an upper bound for the expectation of the average squared gradient norm $\frac{1}{J}\mathbb{E}\left[\sum_{j=0}^{J-1}\left\|\nabla\mathcal{L}(\boldsymbol{\theta}^j)\right\|^2\right]$, to prove the convergence of the proposed AgeSel. We start by presenting the following lemma.

\textbf{Lemma 1:} \textit{Under Assumptions 1-2 and $\eta$ is chosen such that $\eta\leq \frac{1}{8LU}$, there exists a positive constant $c<\frac{1}{2}-15U^2\eta^2 L^2-L\eta (90U^3L^2\eta^2+3U)$, such that}
\begin{align}
    &\mathbb{E}[\mathcal{L}(\boldsymbol{\theta}^{j+1})]\leq \mathcal{L}(\boldsymbol{\theta}^{j})-c\eta U\left\|\nabla \mathcal{L}(\boldsymbol{\theta}^{j})\right\|^2 + \frac{\eta^2 UL}{2SB}\sigma_L^2+Z_1 +L\eta^2Z_2+\frac{2L\eta^2(A^j)^2}{S^2}Z_2-\frac{2L\eta^2A^j}{S}Z_2,
\end{align}
\textit{where we have defined $Z_1=\frac{5U^2\eta^3L^2}{2}\left(\sigma_L^2+6U\sigma_G^2\right)$, $Z_2=15U^3L^2\eta^2(\sigma_L^2+6U\sigma_G^2)+3U^2\sigma_G^2$; and $A^j$ is the cardinality of the set $\mathcal{A}^j$ of workers selected by age at round $j$, i.e., $A^j=\min\{S,S^j\}$.}
\begin{proof}
    The proof can be found in the appendix.
\end{proof}
Lemma 1 depicts the one-step difference of the objective function, from which we can see the impact of the age. Particularly, as $A^j$ grows larger, the variance term $-\frac{2L\eta^2A^j}{S}Z_2$ is reduced, while the variance term $\frac{2L\eta^2(A^j)^2}{S^2}Z_2$ is increased. Therefore, the values of $\tau_{max}$ and $S$, deciding $A^j$ jointly, have a significant impact on training speed, which will be further demonstrated in Section IV. Based on Lemma 1, we are able to arrive at the final convergence result.

\textbf{Theorem 1: } \textit{With the same conditions as in Lemma 1, we have}
\begin{align}
    \frac{1}{J}\mathbb{E}\left[\sum_{j=0}^{J-1}\left\|\nabla\mathcal{L}(\boldsymbol{\theta}^j)\right\|^2\right] \leq \frac{\mathcal{L}(\boldsymbol{\theta}^0)-\mathcal{L}^*}{c\eta U J} + V,\label{theorem1}
\end{align}
\textit{where we have defined the constant $V=\frac{1}{c}\Big[\frac{\eta L}{2SB}\sigma_L^2+\frac{Z_1}{\eta U}+(3\eta L-\frac{2L\eta M}{SR})\frac{Z_2}{U}\Big]$, and $R$ denotes the maximum number of rounds to traverse all the workers in $\mathcal{M}$ due to the age-based mechanism}.

\begin{proof}
The proof can be found in the appendix.
\end{proof}

Theorem 1 states that the proposed AgeSel scheme can achieve a sublinear convergence rate $\mathcal{O}(\frac{1}{J})$ as local SGD with partial worker participation\cite{li2019convergence,yang2020achieving}. The advantage of age-based worker selection is reflected in the expression of $V$, i.e., a smaller $\tau_{max}$ leads to a smaller $R$, and then implies a reduced $V$.

\section{Simulation Results}

In this section, we evaluate the effectiveness of the proposed AgeSel against state-of-the-art schemes including FedAvg\cite{mcmahan2017communication}, OCS\cite{chen2020optimal} and Round Robin (RR) \cite{yang2019scheduling}. Note that  AgeSel with $\tau_{max}$ being large reduces to FedAvg. Furthermore, when $\tau_{max}$ goes to zero, it is obvious that AgeSel is equivalent to RR. Since both $\tau_{max}$ and $S$ determine the value of $A^j$, we will also explore the impact of the hyper-parameter $S$ on the performance of AgeSel.

We start by briefly introducing the three baseline algorithms. For FedAvg, the PS performs weighted selection according to the sizes of the local datasets and we have $\mathcal{M}_D^j=\mathcal{M}_U^j$ with $|\mathcal{M}_D^j|=S$. For OCS, the PS sends the global model to all the workers in $\mathcal{M}$ to perform local computation and only the $S$ workers with larger contribution
are selected, i.e., we have $\mathcal{M}_D^j=\mathcal{M}$ and $|\mathcal{M}_U^j|=S$. For RR, the workers are selected in a circular order with $|\mathcal{M}_D^j|=|\mathcal{M}_U^j|=S$ and the aggregation of the updates is weighted for fair comparison.





\textbf{Simulation Setting.} The dataset $\mathcal{D}$ considered here is the EMNIST dataset. We aim to solve the image classification task with a two-layer fully connected neural network. There are $M=20$ workers in total and the data is heterogeneously distributed among them. Particularly, the samples of the dataset $\mathcal{D}$ is sorted according to its labels and allocated to each worker in order with different sizes. The stepsize $\eta$ is 0.1; the batchsize $B$ is 100 and the number of local updates performed per round $U$ is set to be 5. All the schemes stop training once the test accuracy reaches 80\%. For comparison, the number $S$ of workers selected to upload their local parameters at each round in all three schemes is designated to be $S=5$. Moreover, we set $\tau_{max}=4$ for AgeSel. 

\textbf{AgeSel Outperforms the State-of-Art Schemes in Both Performance Metrics.} The performances of FedAvg, OCS, RR and AgeSel in terms of training rounds and communication cost with 10 Monte Carlo runs are depicted in Fig. 1 and Fig. 2, respectively.  

As shown in Fig. 1, by either considering the weights (i.e., the sizes of datasets) or the ages only, FedAvg and RR require more training rounds than OCS and AgeSel in the presence of data heterogeneity. With larger norms for worker selection, OCS can converge faster. Clearly, AgeSel performs the best by achieving a desirable balance between ages and weights.

\begin{figure}[t!]
\centering
\includegraphics[width=3.6in]{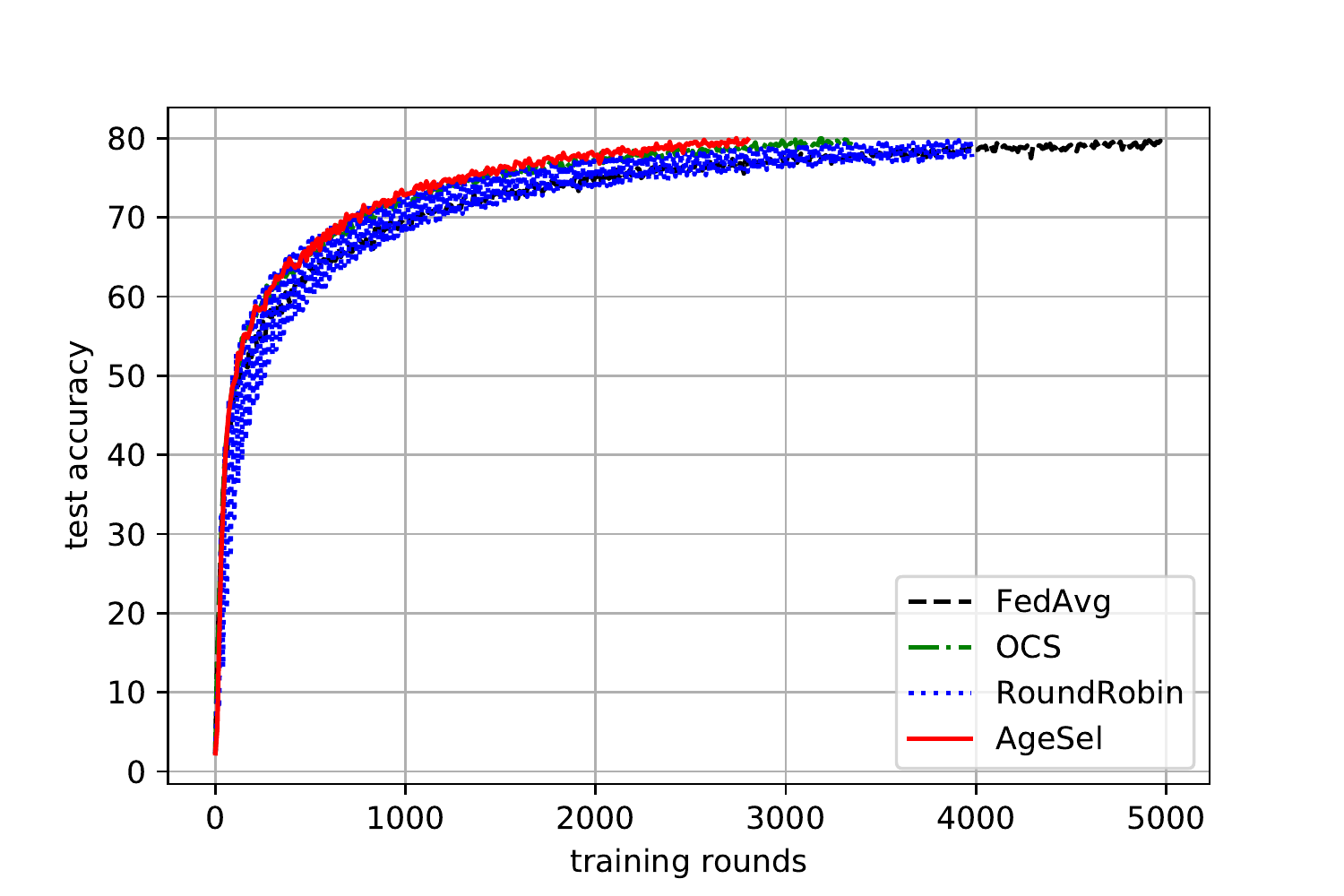}
 \vspace{-6mm}
\caption{Comparison of FedAvg, OCS, RR and AgeSel in terms of training rounds with 10 Monte Carlo runs.}
\vspace{-3mm}
\end{figure}

\begin{figure}[t!]
\centering
\includegraphics[width=3.6in]{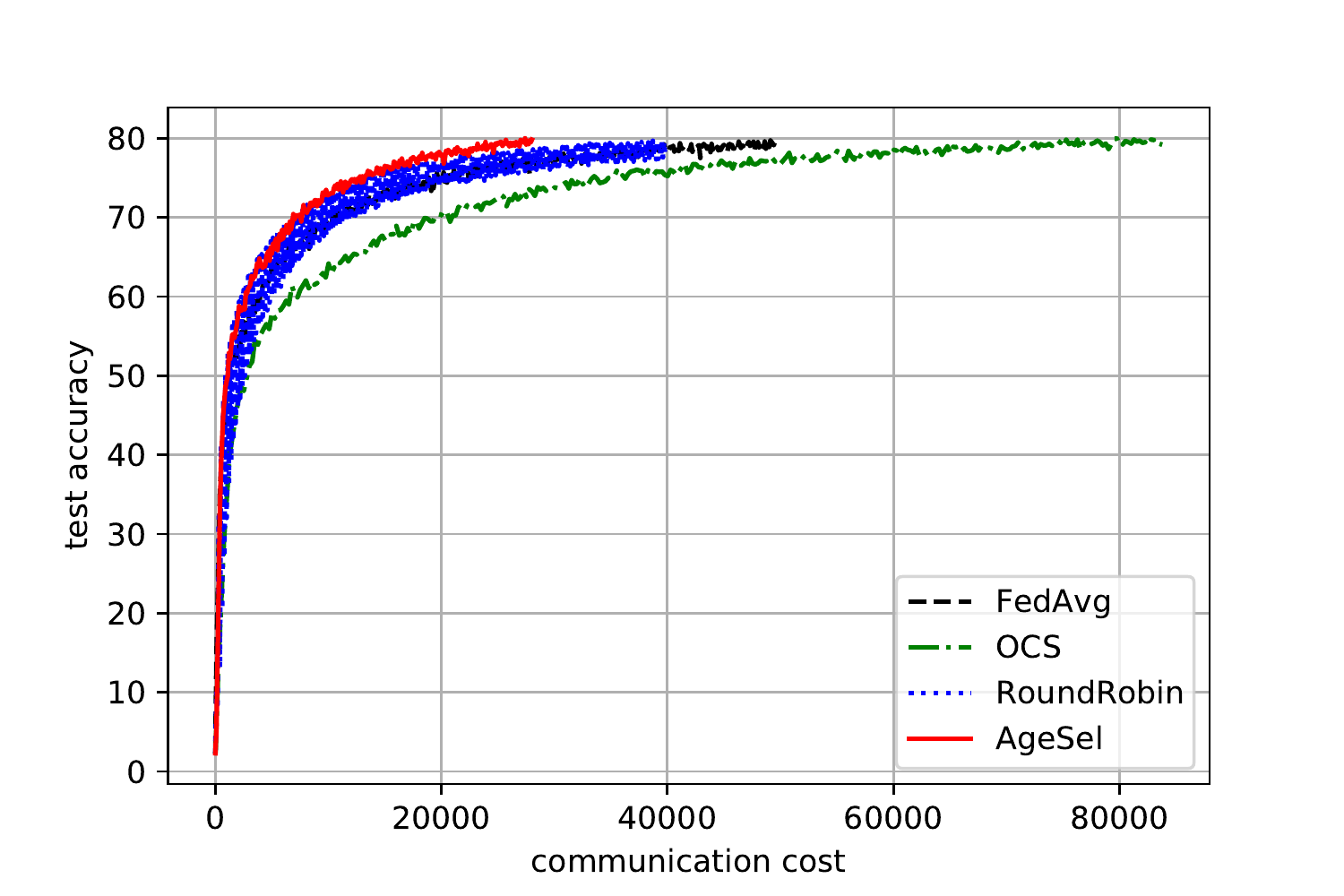}
\vspace{-7mm}
\caption{Comparison of FedAvg, OCS, RR and AgeSel in terms of communication cost with 10 Monte Carlo runs.}
\vspace{-3mm}
\end{figure}

As illustrated in Fig. 2, OCS has the largest communication cost because all the $M$ workers download the global model while $S$ of them are eventually selected. As a result, the reduced training rounds cannot offset the increase of the per-round communication overhead, yielding a larger total cost than the other schemes. Since AgeSel requires fewer training rounds than FedAvg and RR, and they all have the same per-round communication cost, AgeSel achieves better communication efficiency than both FedAvg and RR. Overall, AgeSel is the most communication-efficient selection strategy among all these schemes.



\begin{figure}[t!]
\centering
\includegraphics[width=3.6in]{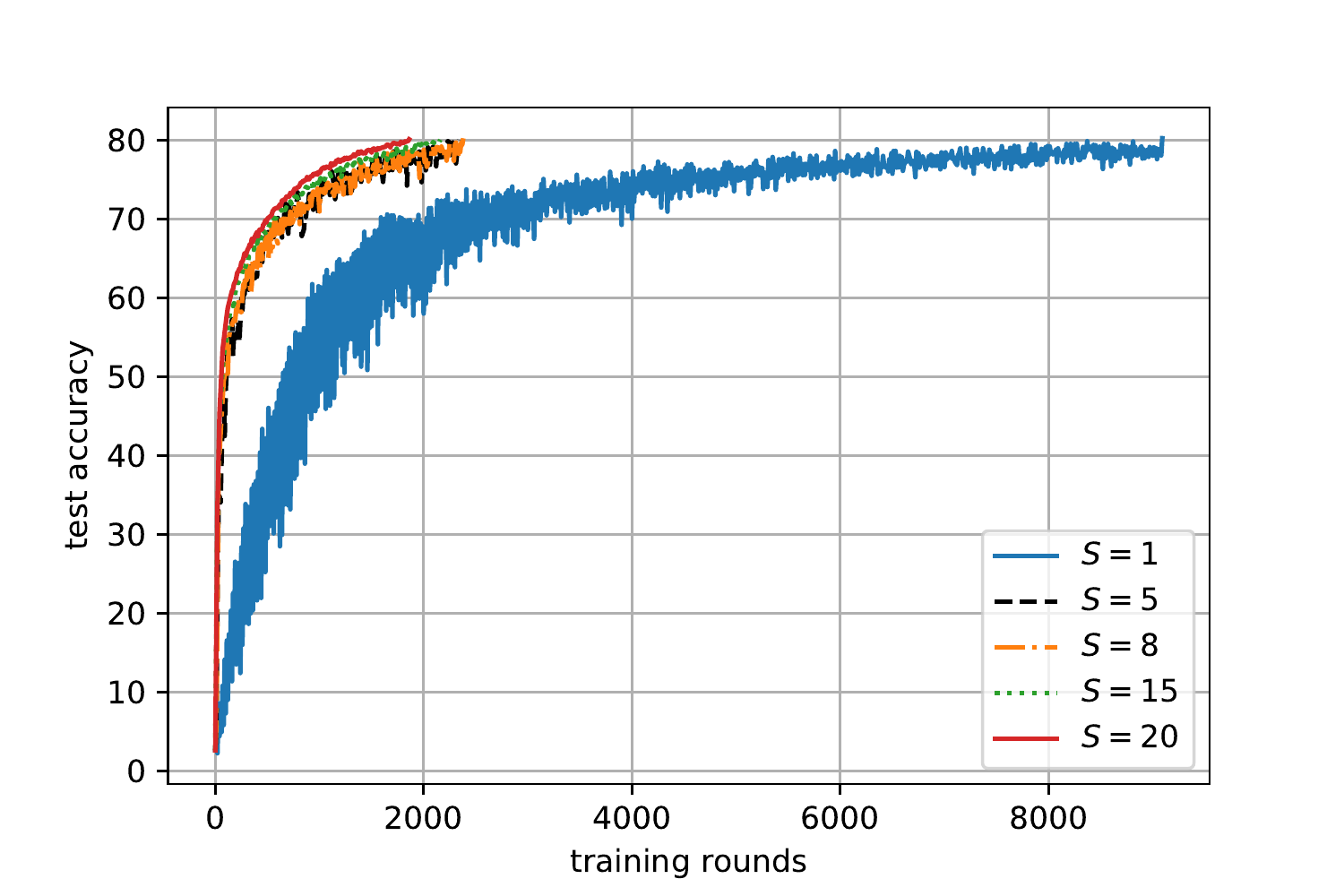}
\vspace{-6mm}
\caption{Number of training rounds of AgeSel with different values of $S$.}
\vspace{-3mm}
\end{figure}

\begin{figure}[t!]
\centering
\includegraphics[width=3.6in]{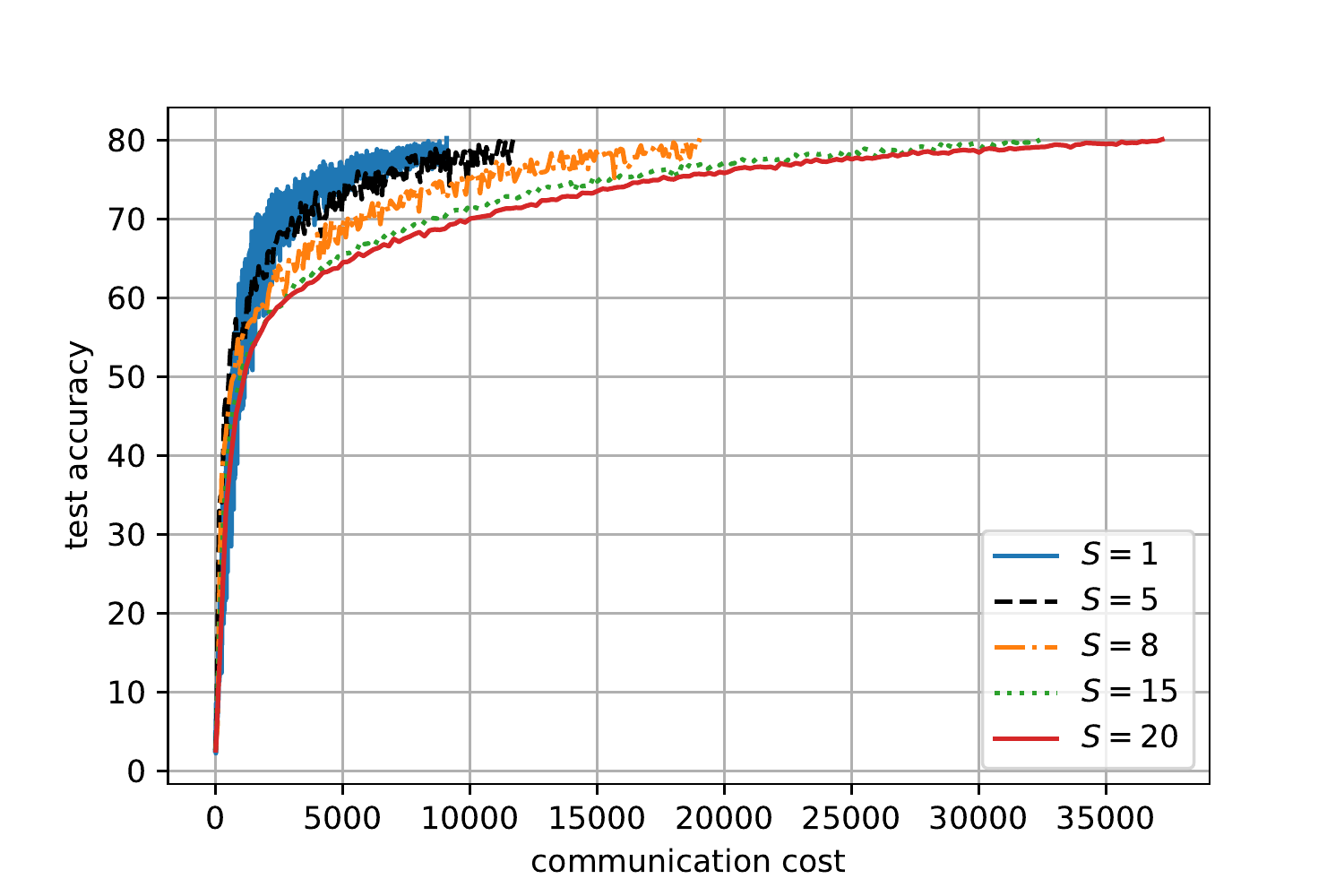}
\vspace{-6mm}
\caption{Communication cost of AgeSel with different values of $S$.}
\vspace{-1mm}
\end{figure}

\textbf{Exploration of $S$.} As illustrated in Fig. 3, when $S$ is larger than some value (e.g., $S=5$), the performance of AgeSel in terms of training round with partial participation becomes quite close to that of full participation (i.e., $S=M=20$). This implies that the age can indeed accelerate the training process. However, as $S$ increases, the per-round communication cost also monotonically increases, as shown in Fig. 4. From Fig. 1$\sim$4, we can see that, with proper values of $S$ and $\tau_{max}$, AgeSel can strike a desirable balance in terms of convergence speed and communication overhead.



\section{Conclusions}

We developed a novel AgeSel strategy to perform adaptive worker selection  for PS-based local SGD under heterogeneous data distribution. The convergence of AgeSel scheme was rigorously proved. Simulation results showed that AgeSel is more communication-efficient and converges faster than state-of-the-art schemes.

Note that AgeSel is compatible with other techniques to jointly improve the overall performance of distributed learning. To name a few, it can be combined with straggler-tolerant techniques to render the learning robust to stragglers; it can also be used together with variance-reduction algorithms to further accelerate the training process. These interesting directions will be pursued in future work.

\section{Appendix}
\subsection{Proof of Lemma 1}

With the $L$-smoothness of the objective function $\mathcal{L}$ in Assumption 1, by taking expectation of all the randomness over $\mathcal{L}(\boldsymbol{\theta}^{j+1})$ we have:
\begin{align}
    \mathbb{E}[\mathcal{L}(\boldsymbol{\theta}^{j+1})]&\leq \mathcal{L}(\boldsymbol{\theta}^{j}) + \left\langle \nabla \mathcal{L}(\boldsymbol{\theta}^{j}), \mathbb{E}[\boldsymbol{\theta}^{j+1}-\boldsymbol{\theta}^j] \right\rangle + \frac{L}{2}\mathbb{E}[\left\|\boldsymbol{\theta}^{j+1}-\boldsymbol{\theta}^j\right\|^2]\notag\\
    &\overset{(a1)}{=}\mathcal{L}(\boldsymbol{\theta}^{j}) + \left\langle \nabla \mathcal{L}(\boldsymbol{\theta}^{j}), \mathbb{E}[g^j+\eta U\nabla \mathcal{L}(\boldsymbol{\theta}^{j})-\eta U\nabla \mathcal{L}(\boldsymbol{\theta}^{j})] \right\rangle + \frac{L}{2}\mathbb{E}[\left\|g^j\right\|^2]\notag\\
    &\overset{(b1)}{=} \mathcal{L}(\boldsymbol{\theta}^{j})-\eta U\left\|\nabla \mathcal{L}(\boldsymbol{\theta}^{j})\right\|^2 + \underbrace{\left\langle \nabla \mathcal{L}(\boldsymbol{\theta}^{j}), \mathbb{E}[g^j+\eta U\nabla \mathcal{L}(\boldsymbol{\theta}^{j})] \right\rangle}_{T_1} + \frac{L}{2}\underbrace{\mathbb{E}[\left\|g^j\right\|^2]}_{T_2},\label{onestep}
\end{align}
where $(a1)$ is due to the definitions $g^j=\frac{1}{S}\sum_{m\in\mathcal{M}_U^j}g_m^j$ where $p_m=N_m/N$ and $g_m^j=-\eta\sum_{u=0}^{U-1}\frac{1}{B}\sum_{b=1}^B\nabla \mathcal{L}(\boldsymbol{\theta}_{m}^{j,u}; z_{m,b}^{j,u})$; and $(b1)$ can be obtained directly from (a1).

We then bound the terms $T_1$ and $T_2$ respectively. For the term $T_1$, with the definition of the weighted global update $\bar{g}^j=\sum_{m\in\mathcal{M}}p_m g_m^j$ and the unbiased global update $\tilde{g}^j=\frac{1}{M}\sum_{m\in\mathcal{M}} g_m^j$, we have 
\begin{align}
    T_1& = \left\langle \nabla \mathcal{L}(\boldsymbol{\theta}^{j}), \mathbb{E}[g^j+\eta U\nabla \mathcal{L}(\boldsymbol{\theta}^{j})] \right\rangle \notag\\
     &\overset{(a2)}{=}\left\langle \nabla \mathcal{L}(\boldsymbol{\theta}^{j}), \mathbb{E}\left[\frac{1}{S}\left(\sum_{m\in\mathcal{M}_U^j\backslash\mathcal{A}^j}g_m^j+\sum_{m\in\mathcal{A}^j}g_m^j\right)+\frac{1}{S}\left((S-A^j)\eta U\nabla \mathcal{L}(\boldsymbol{\theta}^{j})+A^j\eta U\nabla \mathcal{L}(\boldsymbol{\theta}^{j})\right)\right] \right\rangle\notag\\
     &\overset{(b2)}{=}\left\langle \nabla \mathcal{L}(\boldsymbol{\theta}^{j}), \mathbb{E}\left[\frac{S-A^j}{S}\left(\bar{g}^j+\eta U\nabla \mathcal{L}(\boldsymbol{\theta}^{j})\right)\right] \right\rangle + \left\langle \nabla \mathcal{L}(\boldsymbol{\theta}^{j}), \mathbb{E}\left[\frac{A^j}{S}\left(\tilde{g}^j+\eta U\nabla \mathcal{L}(\boldsymbol{\theta}^{j})\right)\right] \right\rangle \label{step1}
\end{align}
where $(a2)$ splits the selected workers at round $j$ into the ones selected by ages in set $\mathcal{A}^j$ with $|\mathcal{A}^j|=A^j=\min\{S,S^j\}$ and the ones selected by weights in set $\mathcal{M}_U^j\backslash\mathcal{A}^j$; $(b2)$ is because the selection by ages is essentially unweighted random selection in expectation and the selection by weights is equivalent to $\bar{g}^j$ in expectation. The two terms in (\ref{step1}) are then bounded separately.

To bound the first term, we have:
\begin{align}
    &\quad\left\langle \nabla \mathcal{L}(\boldsymbol{\theta}^{j}), \mathbb{E}\left[\frac{S-A^j}{S}\left(\bar{g}^j+\eta U\nabla \mathcal{L}(\boldsymbol{\theta}^{j})\right)\right] \right\rangle\notag\\
    & \overset{(a3)}{=} \frac{S-A^j}{S}\left\langle \nabla \mathcal{L}(\boldsymbol{\theta}^{j}), \mathbb{E}\left[-\frac{1}{B}\sum_{m=1}^M\sum_{u=0}^{U-1}\sum_{b=1}^B\eta p_m \nabla\mathcal{L}_m(\boldsymbol{\theta}_{m}^{j,u};z_{m,b}^{j,u}) +\eta U\nabla \mathcal{L}(\boldsymbol{\theta}^{j})\right] \right\rangle \notag\\
    & \overset{(b3)}{=} \frac{S-A^j}{S}\left\langle \nabla \mathcal{L}(\boldsymbol{\theta}^{j}), \mathbb{E}\left[-\sum_{m=1}^M\sum_{u=0}^{U-1}\eta p_m\nabla\mathcal{L}_m(\boldsymbol{\theta}_{m}^{j,u}) +\eta U\sum_{m=1}^M p_m\nabla \mathcal{L}(\boldsymbol{\theta}^{j})\right] \right\rangle \notag\\
    & \overset{(c3)}{=} \frac{S-A^j}{S}\left\langle\sqrt{\eta U} \nabla \mathcal{L}(\boldsymbol{\theta}^{j}), -\frac{\sqrt{\eta}}{\sqrt{U}}\mathbb{E}\left[\sum_{m=1}^M p_m \sum_{u=0}^{U-1}\left(\nabla\mathcal{L}_m(\boldsymbol{\theta}_{m}^{j,u}) - \nabla\mathcal{L}(\boldsymbol{\theta}^j)\right)\right]\right\rangle\notag\\
    & \overset{(d3)}{\leq} \frac{S-A^j}{S}\left(\frac{\eta U}{2} \left\|\nabla \mathcal{L}(\boldsymbol{\theta}^{j})\right\|^2 + \frac{\eta}{2U}\mathbb{E}\left[\left\|\sum_{m=1}^M p_m\sum_{u=0}^{U-1}\left(\nabla\mathcal{L}_m(\boldsymbol{\theta}_{m}^{j,u}) - \nabla\mathcal{L}(\boldsymbol{\theta}^j)\right)\right\|^2\right]\right)\notag\\
    & \overset{(e3)}{\leq} \frac{S-A^j}{S}\left(\frac{\eta U}{2} \left\|\nabla \mathcal{L}(\boldsymbol{\theta}^{j})\right\|^2 + \frac{\eta}{2}\sum_{m=1}^M p_m\sum_{u=0}^{U-1}\mathbb{E}\left[\left\|\left(\nabla\mathcal{L}_m(\boldsymbol{\theta}_{m}^{j,u}) - \nabla\mathcal{L}(\boldsymbol{\theta}^j)\right)\right\|^2\right]\right)\notag\\
    & \overset{(f3)}{\leq} \frac{S-A^j}{S}\left(\frac{\eta U}{2} \left\|\nabla \mathcal{L}(\boldsymbol{\theta}^{j})\right\|^2 + \frac{\eta L^2}{2}\sum_{m=1}^M p_m\sum_{u=0}^{U-1}\mathbb{E}\left[\left\|\boldsymbol{\theta}_{m}^{j,u}-\boldsymbol{\theta}^j\right\|^2\right]\right)\notag\\
    & \overset{(g3)}{\leq} \frac{S-A^j}{S}\left(\eta U\left(\frac{1}{2}+15 U^2\eta^2 L^2\right) \left\|\nabla \mathcal{L}(\boldsymbol{\theta}^{j})\right\|^2 + \frac{5U^2\eta^3L^2}{2}\left(\sigma_L^2+6U\sigma_G^2\right)\right),\label{A1}
\end{align}
where $(a3)$ is derived from the definition of $\bar{g}^j$; $(b3)$ and $(c3)$ come from direct computation; $(d3)$ uses the fact that $\langle\mathbf{x},\mathbf{y}\rangle\leq\frac{1}{2}[\|\mathbf{x}\|^2+\|\mathbf{y}\|^2]$; $(e3)$ is due to Jensen inequality and Cauchy-Schwartz inequality; $(f3)$ follows from Assumption 1; and $(g3)$ is due to the fact that $\sum_{m=1}^M p_m= 1$ and \cite[Lemma 3]{reddi2020adaptive}, which proves that
\begin{align}
\mathbb{E}\left[\left\|\boldsymbol{\theta}_{m}^{j,u}-\boldsymbol{\theta}^j\right\|^2\right]\leq5U\eta^2(\sigma_L^2+6U\sigma_G^2)+30U^2\eta^2 \left\|\nabla \mathcal{L}(\boldsymbol{\theta}^{j})\right\|^2,\label{reddi}
\end{align}
under the condition that $\eta\leq\frac{1}{8LU}$, where $\sigma_L$ and $\sigma_G$ are two constants defined in Assumption 2.

Likewise, the second term in (\ref{step1}) can be bounded as below, with $p_m$ replaced by $\frac{1}{M}$:
\begin{align}
    &\quad\left\langle \nabla \mathcal{L}(\boldsymbol{\theta}^{j}), \mathbb{E}\left[\frac{A^j}{S}\left(\tilde{g}^j+\eta U\nabla \mathcal{L}(\boldsymbol{\theta}^{j})\right)\right] \right\rangle\notag\\
    & \leq \frac{A^j}{S}\left(\eta U\left(\frac{1}{2}+15 U^2\eta^2 L^2\right) \left\|\nabla \mathcal{L}(\boldsymbol{\theta}^{j})\right\|^2 + \frac{5 U^2\eta^3L^2}{2}\left(\sigma_L^2+6U\sigma_G^2\right)\right),\label{A2}
\end{align}

Substituting (\ref{A1}) and (\ref{A2}) into (\ref{step1}), we have
\begin{align}
    T_1&\leq \eta U\left(\frac{1}{2}+15U^2\eta^2L^2\right)\left\|\nabla \mathcal{L}(\boldsymbol{\theta}^{j})\right\|^2 + \frac{5 U^2\eta^3L^2}{2}\left(\sigma_L^2+6U\sigma_G^2\right).\label{T1}
\end{align}

With $\mathbb{I}\{\cdot\}$ denoting the indicator function, and $\mathcal{A}\backslash\mathcal{B}$ denotes the complementary set of set $\mathcal{B}$ in set $\mathcal{A}$, the term $T_2$ can be bounded as
\begin{align}
    T_2 &= \mathbb{E}[\|g^j\|^2]\notag\\
    &\overset{(a4)}{=}\mathbb{E}\left[\left\|\frac{1}{S}\sum_{m\in\mathcal{M}_U^j}g_m^j\right\|^2\right] \notag\\
    &\overset{(b4)}{=}\frac{1}{S^2}\mathbb{E}\left[\left\|\sum_{m=1}^M \mathbb{I}\{m\in\mathcal{M}_U^j\}g_m^j\right\|^2\right]\notag\\
    &\overset{(c4)}{=} \frac{\eta^2}{S^2}\mathbb{E}\left[\left\|\sum_{m=1}^M \mathbb{I}\{m\in\mathcal{M}_U^j\}\sum_{u=0}^{U-1}\left[\frac{1}{B}\sum_{b=1}^B\left(\nabla\mathcal{L}_m(\boldsymbol{\theta}_{m}^{j,u};z_{m,b}^{j,u})-\nabla\mathcal{L}_m(\boldsymbol{\theta}_{m}^{j,u})\right)\right]\right\|^2\right]\notag\\
    \quad&+\frac{\eta^2}{S^2}\mathbb{E}\left[\left\|\sum_{m=1}^M \mathbb{I}\{m\in\mathcal{M}_U^j\}\sum_{u=0}^{U-1}\nabla\mathcal{L}_m(\boldsymbol{\theta}_{m}^{j,u})\right\|^2\right]\notag\\
    &\overset{(d4)}{=} \frac{\eta^2}{S^2B^2}\mathbb{E}\left[\left\|\sum_{m=1}^M \mathbb{I}\{m\in\mathcal{M}_U^j\}\sum_{u=0}^{U-1}\sum_{b=1}^B\left(\nabla\mathcal{L}_m(\boldsymbol{\theta}_{m}^{j,u};z_{m,b}^{j,u})-\nabla\mathcal{L}_m(\boldsymbol{\theta}_{m}^{j,u})\right)\right\|^2\right]\notag\\
    \quad&+\frac{\eta^2}{S^2}\mathbb{E}\left[\left\|\sum_{m=1}^M \mathbb{I}\{m\in\mathcal{M}_U^j\}\sum_{u=0}^{U-1}\nabla\mathcal{L}_m(\boldsymbol{\theta}_{m}^{j,u})\right\|^2\right]\notag\\
    &\overset{(e4)}{=} \frac{\eta^2U}{SB}\sigma_L^2+\frac{\eta^2}{S^2}\mathbb{E}\left[\left\|\sum_{m=1}^M \mathbb{I}\{m\in\mathcal{M}_U^j\}\sum_{u=0}^{U-1}\nabla\mathcal{L}_m(\boldsymbol{\theta}_{m}^{j,u})\right\|^2\right]\notag\\
    &\overset{(f4)}{=} \frac{\eta^2U}{SB}\sigma_L^2+\frac{\eta^2}{S^2}\mathbb{E}\left[\left\|\sum_{m\in\mathcal{A}^j} \sum_{u=0}^{U-1}\nabla\mathcal{L}_m(\boldsymbol{\theta}_{m}^{j,u})+\sum_{m\in\mathcal{M}_U^j\backslash\mathcal{A}^j} \sum_{u=0}^{U-1}\nabla\mathcal{L}_m(\boldsymbol{\theta}_{m}^{j,u})\right\|^2\right]\notag\\
    &\overset{(g4)}{\leq} \frac{\eta^2U}{SB}\sigma_L^2+\frac{2A^j\eta^2}{S^2}\sum_{m\in\mathcal{A}^j}\mathbb{E}\left[\left\| \sum_{u=0}^{U-1}\nabla\mathcal{L}_m(\boldsymbol{\theta}_{m}^{j,u})\right\|^2\right]+\frac{2(S-A^j)\eta^2}{S^2}\sum_{m\in\mathcal{M}_U^j\backslash\mathcal{A}^j} \mathbb{E}\left[\left\|\sum_{u=0}^{U-1}\nabla\mathcal{L}_m(\boldsymbol{\theta}_{m}^{j,u})\right\|^2\right],\label{T_2_temp}
\end{align}
where $(a4)$ is due to the definition of $g^j$; $(b4)$ comes directly from $(a3)$; $(c4)$ follows from the fact that $\mathbb{E}[\|x\|^2]=\mathbb{E}[\|x-\mathbb{E}[x]\|^2+\|\mathbb{E}[x]\|^2]$ and $\mathbb{E}[\nabla\mathcal{L}_m(\boldsymbol{\theta}_{m}^{j,u};z_{m,b}^{j,u})]=\nabla\mathcal{L}_m(\boldsymbol{\theta}_{m}^{j,u})$; $(d4)$ follows from Cauchy-Schwartz inequality; $(e4)$ is due to the fact that $\mathbb{E}[\|x_1+...+x_n\|^2]=\mathbb{E}[\|x_1\|^2+...+\|x_n\|^2]$ if $x_i'$s are independent with zero mean; $(f4)$ is due to the definition that the subset of workers selected by ages at round $j$ is $\mathcal{A}^j$ and $|\mathcal{A}^j|=A^j=\min\{S,S^j\}$; $(g4)$ is due to Cauchy-Schwartz inequality.

Next, we bound the term $\mathbb{E}\left[\left\| \sum_{u=0}^{U-1}\nabla\mathcal{L}_m(\boldsymbol{\theta}_{m}^{j,u})\right\|^2\right]$ in (\ref{T_2_temp}) and (\ref{A1}) as follows:
\begin{align}
    \mathbb{E}\left[\left\| \sum_{u=0}^{U-1}\nabla\mathcal{L}_m(\boldsymbol{\theta}_{m}^{j,u})\right\|^2\right]&=\mathbb{E}\left[\left\| \sum_{u=0}^{U-1}\left(\nabla\mathcal{L}_m(\boldsymbol{\theta}_{m}^{j,u})-\nabla\mathcal{L}_m(\boldsymbol{\theta}^j) +\nabla\mathcal{L}_m(\boldsymbol{\theta}^j)-\nabla\mathcal{L}(\boldsymbol{\theta}^j)+\nabla\mathcal{L}(\boldsymbol{\theta}^j)\right)\right\|^2\right]\notag\\
    &\overset{(a5)}{\leq} 3UL^2\sum_{u=0}^{U-1}\mathbb{E}[\|\boldsymbol{\theta}_{m}^{j,u}-\boldsymbol{\theta}^j\|^2]+3U^2\sigma_G^2+3U^2\|\nabla\mathcal{L}(\boldsymbol{\theta}^j)\|^2\notag\\
    &\overset{(b5)}{\leq} 15U^3L^2\eta^2(\sigma_L^2+6U\sigma_G^2)+(90U^4L^2\eta^2+3U^2)\|\nabla\mathcal{L}(\boldsymbol{\theta}^j)\|^2+3U^2\sigma_G^2\notag\\
    &\overset{(c5)}{=}C_1\|\nabla\mathcal{L}(\boldsymbol{\theta}^j)\|^2+C_2,\label{ti}
\end{align}
where $(a5)$ is due to Cauchy-Schwartz inequality and the bounded variance assumption; $(b5)$ follows from (\ref{reddi}); $(c5)$ is due to the definition that $C_1=90U^4L^2\eta^2+3U^2$ and $C_2=15U^3L^2\eta^2(\sigma_L^2+6U\sigma_G^2)+3U^2\sigma_G^2$.


By substituting the upper bounds (\ref{T1}), (\ref{T_2_temp}) of the terms $T_1$ and $T_2$ in (\ref{onestep}), we readily have:

\begin{align}
    \mathbb{E}[\mathcal{L}(\boldsymbol{\theta}^{j+1})] & \overset{(a6)}{\leq} \mathcal{L}(\boldsymbol{\theta}^{j}) -\eta U\left(\frac{1}{2}-15U^2\eta^2 L^2\right)\left\|\nabla \mathcal{L}(\boldsymbol{\theta}^{j})\right\|^2+\frac{5U^2\eta^3L^2}{2}\left(\sigma_L^2+6U\sigma_G^2\right)+\frac{\eta^2UL}{2SB}\sigma_L^2\notag\\
    \quad & +\frac{L\eta^2A^j}{S^2} \sum_{m\in\mathcal{A}^j}\mathbb{E}\left[\left\|\sum_{u=0}^{U-1}\nabla\mathcal{L}_m(\boldsymbol{\theta}_{m}^{j,u})\right\|^2\right] +\frac{L\eta^2(S-A^j)}{S^2}\sum_{m\in\mathcal{M}_U^j\backslash\mathcal{A}^j}\mathbb{E}\left[\left\| \sum_{u=0}^{U-1}\nabla\mathcal{L}_m(\boldsymbol{\theta}_{m}^{j,u})\right\|^2\right]\notag\\
    &\overset{(b6)}{\leq}\mathcal{L}(\boldsymbol{\theta}^{j}) -\eta U\left(\frac{1}{2}-15U^2\eta^2 L^2\right)\left\|\nabla \mathcal{L}(\boldsymbol{\theta}^{j})\right\|^2+\frac{5U^2\eta^3L^2}{2}\left(\sigma_L^2+6U\sigma_G^2\right)+\frac{\eta^2UL}{2SB}\sigma_L^2\notag\\
    &+\frac{L\eta^2\left((A^j)^2+(S-A^j)^2\right)}{S^2}\left(C_1\|\nabla\mathcal{L}(\boldsymbol{\theta}^j)\|^2+C_2\right)\notag\\
    &\overset{(c6)}{\leq}\mathcal{L}(\boldsymbol{\theta}^{j})-\eta U\left(\frac{1}{2}-15U^2\eta^2 L^2-\frac{L\eta\left(2(A^j)^2-2SA^j+S^2\right)}{S^2U}C_1\right)\left\|\nabla \mathcal{L}(\boldsymbol{\theta}^{j})\right\|^2\notag\\
    \quad & +\frac{5U^2\eta^3L^2}{2}\left(\sigma_L^2+6U\sigma_G^2\right)+\frac{\eta^2UL}{2SB}\sigma_L^2+\frac{L\eta^2\left(2(A^j)^2-2SA^j+S^2\right)}{S^2}C_2\notag\\
    &\overset{(d6)}{\leq}\mathcal{L}(\boldsymbol{\theta}^{j})-\eta U\left(\frac{1}{2}-15U^2\eta^2 L^2-\frac{L\eta}{U}C_1\right)\left\|\nabla \mathcal{L}(\boldsymbol{\theta}^{j})\right\|^2\notag\\
    \quad & +\frac{5U^2\eta^3L^2}{2}\left(\sigma_L^2+6U\sigma_G^2\right)+\frac{\eta^2UL}{2SB}\sigma_L^2+\frac{L\eta^2\left(2(A^j)^2-2SA^j+S^2\right)}{S^2}C_2\notag\\
    &\overset{(e6)}{\leq}\mathcal{L}(\boldsymbol{\theta}^{j})-c\eta U\left\|\nabla \mathcal{L}(\boldsymbol{\theta}^{j})\right\|^2 +\frac{5U^2\eta^3L^2}{2}\left(\sigma_L^2+6U\sigma_G^2\right)+\frac{\eta^2UL}{2SB}\sigma_L^2+L\eta^2C_2+\frac{2L\eta^2(A^j)^2}{S^2}C_2-\frac{2L\eta^2A^j}{S}C_2,\label{onestep_final}
\end{align}
where $(a6)$ comes from direct substitution; $(b6)$ uses the result in (\ref{ti}); $(c6)$ follows from direct computation; $(d6)$ uses the fact that $0\leq A^j\leq S$; and $(e6)$ follows from the fact that there exists a constant $c$ such that $0<c<\frac{1}{2}-15U^2\eta^2 L^2-L\eta (90U^3L^2\eta^2+3U)$. The proof of Lemma 1 is then complete.

\subsection{Proof of Theorem 1}
With Lemma 1, we have
\begin{align}
    &\mathbb{E}[\mathcal{L}(\boldsymbol{\theta}^{j+1})]\overset{(a8)}{\leq}\mathcal{L}(\boldsymbol{\theta}^{j})-c\eta U\left\|\nabla \mathcal{L}(\boldsymbol{\theta}^{j})\right\|^2 +\frac{5U^2\eta^3L^2}{2}\left(\sigma_L^2+6U\sigma_G^2\right)+\frac{L\eta^2(2(A^j)^2+S^2)}{S^2}-\frac{2L\eta^2A^j}{S}C_2\notag\\
    &\overset{(b8)}{\leq}\mathcal{L}(\boldsymbol{\theta}^{j})-c\eta U\left\|\nabla \mathcal{L}(\boldsymbol{\theta}^{j})\right\|^2 +\frac{5U^2\eta^3L^2}{2}\left(\sigma_L^2+6U\sigma_G^2\right)+\frac{\eta^2UL}{2SB}\sigma_L^2+3L\eta^2C_2-\frac{2L\eta^2A^j}{S}C_2,\label{lemmatemp}
\end{align}
where $(a8)$ is from (\ref{onestep_final}); and $(b8)$ uses the fact that $0\leq A^j\leq S$.

With the age-based mechanism, we denote the minimum number of rounds to traverse all the workers in $\mathcal{M}$ as $R$, i.e., we have $A^j+...+A^{j+R-1}\geq M$. By rearranging the terms in (\ref{lemmatemp}) and summing from $j=0,...,R-1$, we can have:
\begin{align}
    \frac{1}{R}\mathbb{E}\left[\sum_{j=0}^{R-1}\left\|\nabla\mathcal{L}(\boldsymbol{\theta}^j)\right\|^2\right] \leq \frac{\mathcal{L}(\boldsymbol{\theta}^0)-\mathcal{L}(\boldsymbol{\theta}^R)}{c\eta U R} + V_1,\label{theo1_a}
\end{align}
where $V_1=\frac{1}{c\eta U}(\frac{\eta^2UL}{2SB}\sigma_L^2 + \frac{5U^2\eta^3L^2}{2}\left(\sigma_L^2+6U\sigma_G^2\right)+(3L\eta^2-\frac{2L\eta^2M}{SR})C_2)$.

Thus, when $J$ is a multiple of $R$, we can then readily write
\begin{align}
    \frac{1}{J}\mathbb{E}\left[\sum_{j=0}^{J-1}\left\|\nabla\mathcal{L}(\boldsymbol{\theta}^j)\right\|^2\right] \leq \frac{\mathcal{L}(\boldsymbol{\theta}^0)-\mathcal{L}^*}{c\eta U J} + V,\label{theo1}
\end{align}
where we used Assumption 1 and $V=\frac{1}{c}[\frac{\eta L}{2SB}\sigma_L^2+\frac{5U\eta^2L^2}{2}\left(\sigma_L^2+6U\sigma_G^2\right)+(3\eta L-\frac{2L\eta M}{SR})(15U^2L^2\eta^2(\sigma_L^2+6U\sigma_G^2)+3U\sigma_G^2)]$. The proof is then complete. 

\end{document}